\documentclass[twocolumn,tighten,longbib]{aastex7}
\usepackage{graphicx}
\usepackage{amsmath}
\usepackage{scalerel}
\usepackage{apjfonts}
\usepackage{tikz}
\usetikzlibrary{svg.path}

\definecolor{orcidlogocol}{HTML}{A6CE39}
\tikzset{
  orcidlogo/.pic={
    \fill[orcidlogocol] svg{M256,128c0,70.7-57.3,128-128,128C57.3,256,0,198.7,0,128C0,57.3,57.3,0,128,0C198.7,0,256,57.3,256,128z};
    \fill[white] svg{M86.3,186.2H70.9V79.1h15.4v48.4V186.2z}
                 svg{M108.9,79.1h41.6c39.6,0,57,28.3,57,53.6c0,27.5-21.5,53.6-56.8,53.6h-41.8V79.1z M124.3,172.4h24.5c34.9,0,42.9-26.5,42.9-39.7c0-21.5-13.7-39.7-43.7-39.7h-23.7V172.4z}
                 svg{M88.7,56.8c0,5.5-4.5,10.1-10.1,10.1c-5.6,0-10.1-4.6-10.1-10.1c0-5.6,4.5-10.1,10.1-10.1C84.2,46.7,88.7,51.3,88.7,56.8z};
  }
}

\newcommand\orcidicon[1]{\href{https://orcid.org/#1}{\mbox{\scalerel*{
\begin{tikzpicture}[yscale=-1,transform shape]
\pic{orcidlogo};
\end{tikzpicture}
}{|}}}}

\newcommand{\um}{$\mu$m}
\newcommand{\cotwo}{CO$_{2}$}
\newcommand{\water}{H$_{2}$O}
\newcommand{\methanol}{CH$_{3}$OH}

\usepackage{comment}
\usepackage{array,multirow,color}
\usepackage{xcolor}

\begin{document}

\title{JWST Spectroscopy of a Blue Binary Cold Classical Kuiper Belt Object}


\author[0000-0001-9665-8429]{Ian~Wong}
\affiliation{Space Telescope Science Institute, 3700 San Martin Drive, Baltimore, MD 21218, USA}
\email{iwong@stsci.edu}

\author[0000-0002-6117-0164]{Bryan~J.~Holler}
\affiliation{Space Telescope Science Institute, 3700 San Martin Drive, Baltimore, MD 21218, USA}
\email{bholler@stsci.edu}

\author[0000-0001-6680-6558]{Wesley~C.~Fraser}
\affiliation{National Research Council of Canada, Herzberg Astronomy and Astrophysics Research Centre, 5071 W. Saanich Road, Victoria, BC V9E 2E7, Canada}
\affiliation{Department of Physics and Astronomy, University of Victoria, Elliott Building, 3800 Finnerty Road, Victoria, BC V8P 5C2, Canada}
\email{wesley.fraser@nrc-cnrc.gc.ca}

\author[0000-0002-8255-0545]{Michael~E.~Brown}
\affiliation{Division of Geological and Planetary Sciences, California Institute of Technology, Pasadena, CA 91125, USA}
\email{mbrown@caltech.edu}

\begin{abstract}
We present observations of two binary systems within the cold classical region of the Kuiper belt---2001~XR254 and 2016~BP81---obtained with the JWST Near-Infrared Spectrograph. The measured reflectance spectrum of 2001~XR254 is characteristic of the red cold classicals, with strong features due to carbon dioxide, carbon monoxide, and methanol ices. In contrast, 2016~BP81 is a blue binary, with a spectrum that shows prominent water-ice absorption bands. The two components of the 2016~BP81 binary display identical spectral profiles, consistent with coeval formation from gravitational collapse. Through qualitative and quantitative comparisons of Kuiper belt objects (KBOs) with similar spectral profiles dominated by water-ice features, we identify a small cluster, including 2016~BP81, that differs in systematic ways from the rest of the population. The relatively deep carbon dioxide ice absorption bands, different water-ice band-depth ratios, and possibly enhanced signatures of aliphatic organics suggest that objects in this cluster may have originated in a distinct formation environment from the other water-type KBOs. The implications of our findings are discussed within the context of recent models of Kuiper belt formation and evolution.
\end{abstract}

\section{Introduction}
\label{sec:intro}
\setcounter{footnote}{0}

The vast population of icy bodies in the Kuiper belt traces the initial compositional landscape and subsequent dynamical processes in the outer protoplanetary disk and holds the key to understanding many fundamental aspects of early solar system history. As the detailed orbital architecture of the Kuiper belt has come into focus in the past two decades, a surprisingly complex dynamical picture has emerged \citep[e.g.,][]{gladman2008,kavelaars2008,brown2012,kavelaars2020}, which has in turn motivated new theories of solar system evolution. Recent models attribute the broad eccentricity and inclination distribution of Kuiper belt objects (KBOs) to outward migration of Neptune that occurred shortly after the end of planet formation \citep{morbidelli2020,gladman2021}. Numerical simulations have successfully reproduced most dynamical features of the present-day Kuiper belt by invoking a gradual expansion of Neptune's orbit, which led to the disruption of the primordial disk of icy planetesimals that was situated beyond 30~au \citep[e.g.,][]{levison2008,nesvorny2012,nesvorny2016}.

Contemporaneously, a concerted effort has been levied toward characterizing the surface properties of KBOs. Ground- and space-based multiband photometry and low-resolution spectroscopy uncovered a stark color bimodality at both visible and near-infrared wavelengths among objects fainter than an absolute magnitude of $H\sim6$~mag, i.e., smaller than $\sim$300--400~km \citep[e.g.,][]{fraser2012,peixinho2012,peixinho2015,tegler2016,wong2017,schwamb2019,fraser2023}. This bimodality separates KBOs into two subpopulations that we refer to here as less red and red.\footnote{Other works in the literature apply different binary color designations, e.g., gray/red and red/very red.} The color bimodality
suggests that a sharp gradient in surface composition must have existed somewhere within the primordial trans-Neptunian disk, possibly caused by differential sublimation and irradiation chemistry that occurred on two sides of a volatile ice retention line \citep{schaller2007,brown2011,wong2016}.

The advent of JWST has provided the means to place the attested diversity of KBO colors within a more concrete compositional framework. Near-infrared observations of dozens of KBOs in the range $H \sim 3.5-8.5$~mag have revealed three distinct compositional classes \citep{depra2025,licandro2025,pinillaalonso2025}: (1) objects with deep water-ice (\water) bands and less-red visible spectral slopes that also show minor absorptions due to carbon dioxide (\cotwo) ice; (2) objects with colors consistent with the red visible color subpopulation that display an exceptionally deep 4.27~\um~\cotwo~primary band, evidence of carbon monoxide (CO) ice, and strong organic features produced from the irradiation of aliphatic hydrocarbons (e.g., methane, ethane); (3) a separate class of red-colored KBOs with a distinct near-infrared continuum shape that additionally show characteristic features of methanol (\methanol). In the following, we adopt the recent taxonomic nomenclature proposed in \citet{holler2025rnaas} and refer to these three compositional classes as \water-type, \cotwo-type, and organics-type, respectively.\footnote{These three compositional classes are designated by the respective monikers ``bowl,'' `double-dip,'' and ``cliff'' in \citet{pinillaalonso2025}.} Extending the aforementioned volatile ice retention hypothesis to the three types of KBOs implies the existence of two compositional gradients in the surface chemistry of planetesimals within the primordial Kuiper belt region. Based on the observed compositional inventories, the \water-type KBOs, which show the fewest volatile ice signatures on their surfaces, are expected to have formed in the innermost region of the disk, with the \cotwo- and organics-type objects accreting at progressively larger heliocentric distances \citep{pinillaalonso2025}.

The so-called cold classicals---KBOs on relatively low-inclination, low-eccentricity orbits that lie between the 3:2 and 2:1 mean motion resonances with Neptune at 42--47~au---challenge our current understanding of Kuiper belt formation and evolution. This population is notable for containing a high fraction ($\sim$30\%) of near-equal-mass binaries, including many with wide separations \citep{noll2020}. Dynamical studies have demonstrated that a majority of these binary systems would have been lost had they been scattered into their present-day orbits during the outward migration of Neptune \citep{parker2010,nesvorny2019a}. Instead, cold classicals may have formed in situ and therefore comprise a unique representative collection of planetesimals from the outermost reaches of the protoplanetary disk \citep{nesvorny2022}.

Within the aforementioned compositional gradient model, one would expect in situ formation of cold classicals to exclusively produce red objects. While red objects indeed dominate the cold classical population, observations within the last decade have revealed a significant admixture of objects in this region with bluer visible spectral slopes; moreover, these bluer objects are predominantly binaries \citep{fraser2017,fraser2021}. These objects are commonly referred to as blue binaries in the literature, though we note that their colors are broadly consistent with the less-red KBO subpopulation. The blue binaries are unlikely to have formed within the inner portion of the primordial trans-Neptunian planetesimal disk, given the fragility of widely separated binaries to significant dynamical excitation. Instead, it has been proposed that all blue binaries formed as binary pairs immediately inward of the present-day cold classical region, between 38 and 42~au; subsequent slow migration of Neptune's 2:1 mean motion resonance could have gently pushed these objects out into their current orbits, consistent with the somewhat inflated inclination distribution of blue binaries when compared to the redder cold classicals \citep{fraser2017,fraser2021}. 

However, recent attempts to simulate this evolutionary pathway have reached an impasse \citep[e.g.,][]{nesvorny2020,nesvorny2022}. While a color transition immediately inward of the current cold classical region readily produces the blue binaries, it also predicts a significantly larger proportion of less-red objects throughout the inner trans-Neptunian planetesimal disk, particularly below 30~au, where the vast majority of the primordial mass is expected to be concentrated. As a result of Neptune’s migration, a large number of less-red singletons would have been scattered outward and implanted into the cold classical population, inconsistent with the observed preponderance of binaries. A drastic reformulation of the initial conditions in the trans-Neptunian region and the specific trajectory of Neptune's migration would be needed to resolve this quandary. An alternative explanation posits that all cold classicals (including the blue binaries) formed in situ, with the color distinction being a consequence of the timing of formation relative to the dispersion of the gas disk \citep{nesvorny2022}.

\begin{deluxetable*}{lccccccccccccl}
\tablewidth{0pc}
\setlength{\tabcolsep}{3pt}
\tabletypesize{\footnotesize}
\tablecaption{
    Targets and Observation Details
    \label{tab:targets}
}

\tablehead{Target & $a$ (au) & $e$ & $i$ ($^{\circ}$) & $D_1$, $D_2$ (km) & $d_{\mathrm{sep}}$ (km) & $s$ (\%/100~nm) & $t_{\mathrm{start}}$ (UT) & $r$ (au) & $\Delta$ (au) & $\alpha$ ($^{\circ}$) & $V$ (mag) & $t_{\mathrm{exp}}$ (s) & Comments
}
\startdata
2001~XR254 & 42.9 & 0.035 & 1.2 & 172, 140 & 9310 & $15.2\pm4.4$ & 2024 Nov 21 05:13:26 & 43.79 & 43.45 & 1.23 & 22.3 & 2976 & \\
2016~BP81 & 43.7 & 0.074 & 4.2 & 188, 170 & 11300 & $7.7\pm1.7$ & 2025 Jan 18 15:52:09 & 43.29 & 43.27 & 1.31 & 22.6 & 3793 & pointing offset \\
2003~HG57 & 43.8 & 0.038 & 2.1 & 156, 156 & 13200 & $10.2\pm3.4$ & 2025 Feb 15 15:23:07 & 42.45 & 42.31 & 1.33 & 22.8 & 5019 & target missed \\
\enddata
\textbf{Note.} $a$: orbital semimajor axis; $e$: orbital eccentricity; $i$: orbital inclination; $D_1$, $D_2$: diameters of the primary and secondary components; $d_{\mathrm{sep}}$: mutual separation distance; $s$: visible spectral slope; $t_{\mathrm{start}}$: observation start time; $r$: heliocentric distance at time of observation; $\Delta$: distance from JWST at time of observation; $\alpha$: phase angle of observation; $V$: visible apparent magnitude at time of observation; $t_{\mathrm{exp}}$: total exposure time across all four dithers. The orbital parameters and viewing geometry are computed by JPL Horizons. The visible spectral slope measurements are tabulated in \citet{fraser2021}. The sizes and mutual separation of the binary components are taken from \citet{nesvorny2022}.
\vspace{-0.5cm}
\end{deluxetable*}

It is evident that blue binaries are a particularly diagnostic population for validating and refining our fundamental understanding of the Kuiper belt. Obtaining a detailed picture of their surface chemistry and comparing it to the compositional classes previously identified via JWST spectroscopy has the potential to resolve the ongoing conundrum concerning the provenance of blue binaries within the cold classical region. To this end, we obtained observations of several cold classical binaries as part of Cycle 3 JWST Program \#5940. Section~\ref{sec:obs} describes the target selection, observing methods, and data reduction and presents the measured reflectance spectra. Comparative characterization of the spectra and discussion of their implications are provided in Section~\ref{sec:disc}.

\section{Observations and Data Analysis}
\label{sec:obs}

\subsection{Observation Design}
The observations were carried out using the integral field unit (IFU) on the Near-Infrared Spectrograph (NIRSpec; \citealt{jakobsen2022,boker2023}). A set of four dithered exposures was collected for each target using the PRISM grating, which covers the 0.6--5.3~\um\ wavelength range at an average spectral resolution of $\Delta\lambda/\lambda \sim100$. To reduce detector read noise, all exposures were read out with the NRSIRS2RAPID method \citep{moseley2010,rauscher2012}. Due to the large uncertainties in the binary orbit solutions for our targets, no timing constraints were stipulated to schedule the observations to coincide with maximum component separation.

\subsection{Target Selection}
When selecting targets for observation as part of our Cycle 3 JWST program, we consulted the list of cold classicals with visible spectral slope measurements published in \citet{fraser2021}. Restricting our search to confirmed binaries and adopting the spectral slope value of $s = 17\% / 100$~nm as the threshold dividing the less-red and red color subpopulations, we chose the three brightest blue binaries according to their minimum apparent magnitude during the Cycle 3 observing windows. The selected targets were 2001~XR254, 2003~HG57, and 2016~BP81, which have measured spectral slopes of $15.2\pm4.4$, $10.2\pm3.4$, and $7.7\pm1.7$, respectively, in units of \%/100~nm. Table~\ref{tab:targets} provides additional information about the targets, including component diameters, mutual separations, and the details of our observations.

As part of the proposal preparation process, the ephemerides of each target were ingested into the Astronomer's Proposal Tool (APT) via a query of the JPL Horizons service. An important caveat of this functionality is that the ephemerides ingestion is a one-time action and not a dynamic link. As such, changes in the ephemerides from subsequent updates to the JPL Horizons orbital solutions are not automatically transferred to the APT. Our program was initially designed during the Cycle 2 call for proposals, and we neglected to reingest the latest ephemerides prior to the scheduled observations. This oversight proved to be critical for our observations of 2003~HG57, whose predicted position changed drastically in the time since the initial ingest. As a result, the target did not fall within the field of view (FOV) of the NIRSpec IFU. The ephemerides for 2016~BP81 also evolved somewhat, but not enough to cause a failed observation, albeit the location of the target within the FOV was offset from the nominal pointing position. 

\subsection{Data Reduction}\label{subsec:extract}

Data processing and spectral extraction were handled by the \texttt{jwstspec} pipeline \citep{jwstspec}---a dedicated tool for customized end-to-end reduction of JWST spectroscopic datasets that was developed primarily for observations of small bodies in the solar system. This pipeline has been employed in numerous previous analyses of NIRSpec observations \citep[e.g.,][]{emery2024,grundy2024,wong2024}, and a full description of the overall pipeline workflow can be found in those publications. Beginning with the uncalibrated data, \texttt{jwstspec} passed through the first two stages of Version 1.17.1 of the official JWST pipeline \citep{jwst}, with necessary reference files drawn from context \texttt{jwst\_1322.pmap} of the JWST Calibration Reference Data System. All pipeline parameters were left at their default values, except for the \texttt{nsclean} step in the second stage, which was manually activated to correct for residual read noise on the detector. The result of the pipeline processing was bias-corrected, flat-fielded, spatially rectified, wavelength-calibrated, and flux-calibrated IFU data cubes for each dithered exposure. 

\begin{figure}
    \includegraphics[width=\columnwidth]{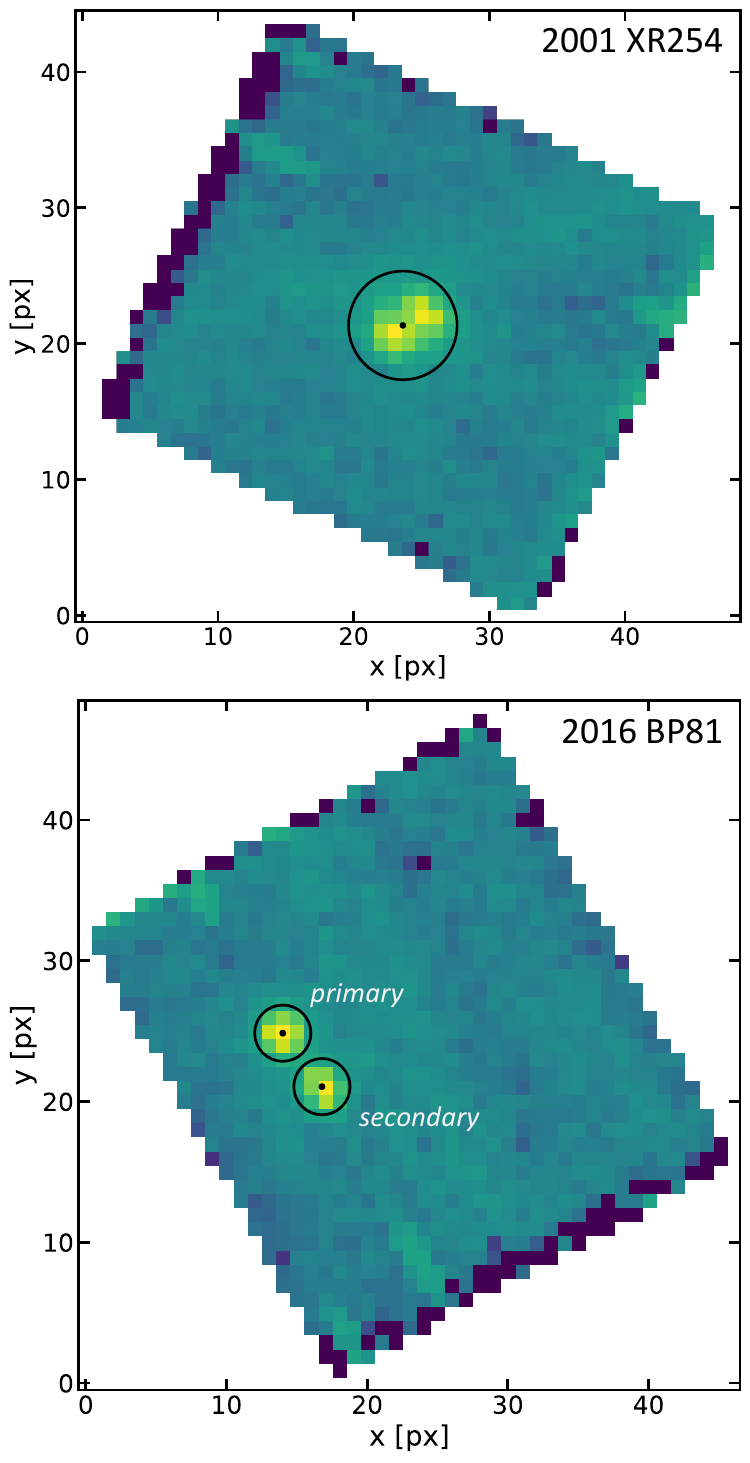}
    \caption{Median-averaged IFU image slices from the first dithered exposures of 2001~XR254 and 2016~BP81. A logarithmic stretch has been applied to the flux scaling to accentuate the PSF shapes and background variations. The black points and circles indicate the measured source centroid positions and spectral extraction apertures. The primary and secondary components of 2016~BP81 are well separated and labeled.}
    \label{fig:images}
\end{figure}

To compute the centroid position of the target within the FOV, the IFU data cubes were collapsed along the wavelength axis to produce a median image of the $3'' \times 3''$ FOV. Figure~\ref{fig:images} shows the respective median images from the first dithered exposures of 2001~XR254 and 2016~BP81. The binary nature of the targets is clearly evident. In the case of 2016~BP81, the observations took place fortuitously near maximum mutual separation, and the two components are individually resolved at the $0\overset{''}{.}1$ pixel scale of the NIRSpec IFU, allowing us to extract the spectrum of each component individually with negligible contamination. The slightly offset pointing for 2016~BP81 is also apparent---a consequence of the deprecated ephemerides. The \texttt{jwstspec} pipeline uses an iterative two-dimensional Gaussian fit to determine the centroid position from the median image. For 2016~BP81, each component was masked when centering the spectral extraction aperture of the other. In the case of 2001~XR254, the point-spread functions (PSFs) of the two components are blended, and we did not attempt to separate the two components when computing the centroid. Nevertheless, even with a single-Gaussian fit, the measured centroid is reasonable, lying intermediate between the two components, as shown in Figure~\ref{fig:images}.

The \texttt{jwstspec} pipeline offers several different spectral extraction techniques. For the results presented in this paper, we utilized the empirical PSF fitting method, which is optimized for faint point-source targets and yields substantial improvements in precision over standard circular aperture extraction \citep[e.g.,][]{wong2024,pinillaalonso2025}. In short, a moving median window of a specified width is applied along the wavelength axis to generate background-subtracted template PSFs, which are then normalized and fit to the central wavelength slices within the windows to obtain the best-fit multiplicative scaling factors for the templates. The fluxes are then measured from the scaled template PSFs within a user-defined circular aperture. See the referenced papers for full details on this method. For our observations, we set the background region as all pixels outside of a $21\times21$ box centered on the measured centroid. The moving median window width was set to 51 (i.e., from $-25$ to $+25$ along the wavelength axis), though adjusting this value by up to $15$ did not substantively change the resultant spectra.  Uncertainties on the fluxes were estimated by weighting the error arrays with the peak-normalized PSF template model and adding in quadrature the uncertainties of all pixels within the extraction aperture.


For 2001~XR254, we selected a 4-pixel-radius circular aperture for extracting the spectra, which envelops both components. For both the primary and secondary components of the 2016~BP81 system, we used a smaller 2-pixel-radius aperture to avoid contamination. The location of the extraction apertures are shown in Figure~\ref{fig:images}.

The spectra extracted from the set of four dithered exposures were passed through a 21~pixel-wide moving median filter to remove $5\sigma$ outliers. The combined irradiance spectrum of each target was obtained by a simple mean of the four individual dithered spectra. Due to low instrument throughput and poor data quality at the extreme ends of the wavelength range, we trimmed the spectra to 0.7--5.1~\um.

\subsection{Accounting for Sensitivity Variations}

We used NIRSpec observations of the G-type solar analog star SNAP-2 when deriving the reflectance spectra of our targets. By dividing the irradiance spectrum of an object by that of the star, processed in an analogous manner and extracted from an aperture with the same equivalent encircled flux, we self-consistently corrected for the wavelength-dependent flux losses outside of the extraction apertures while also removing any common-mode instrumental systematics that are present in the data. In the case of the blended binary 2001~XR254, our choice of the 4-pixel-radius aperture is roughly equivalent to two 3-pixel-radius apertures centered on the two components in terms of encircled flux, and we therefore used a 3-pixel-radius aperture when extracting the stellar spectrum. For 2016~BP81, the stellar spectrum was extracted with the same 2-pixel-radius aperture as the target spectrum.


The offset of 2016~BP81 relative to the nominal pointing region in the center of the FOV is almost $1''$, which is larger than the $\sim$$0\overset{''}{.}4$ range subtended by the 4-point dither pattern. This raises the possibility of systematic sensitivity variations across the FOV that could bias the shape of the resultant reflectance spectrum when incorporating stellar observations in which the source was centered in a different region of the FOV. To investigate this issue, we considered two observations of SNAP-2, one that employed a 20-point dither pattern (Program \#1128, Observation 2) and another that used the 4-point nod pattern (Program \#4498, Observation 19).\footnote{For more on NIRSpec IFU dither and nod patterns, see \url{https://jwst-docs.stsci.edu/jwst-near-infrared-spectrograph/nirspec-operations/nirspec-dithers-and-nods/nirspec-ifu-dither-and-nod-patterns}.} In the former case, the spread in pointing positions is similar to the 4-point dither pattern that was selected for our KBO targets, while in the latter, the internod separations are much larger, with the star placed at the vertices of a $\sim$$1\overset{''}{.}6\times1\overset{''}{.}6$ box centered on the middle of the FOV.

\begin{figure}[t!]
    \includegraphics[width=\columnwidth]{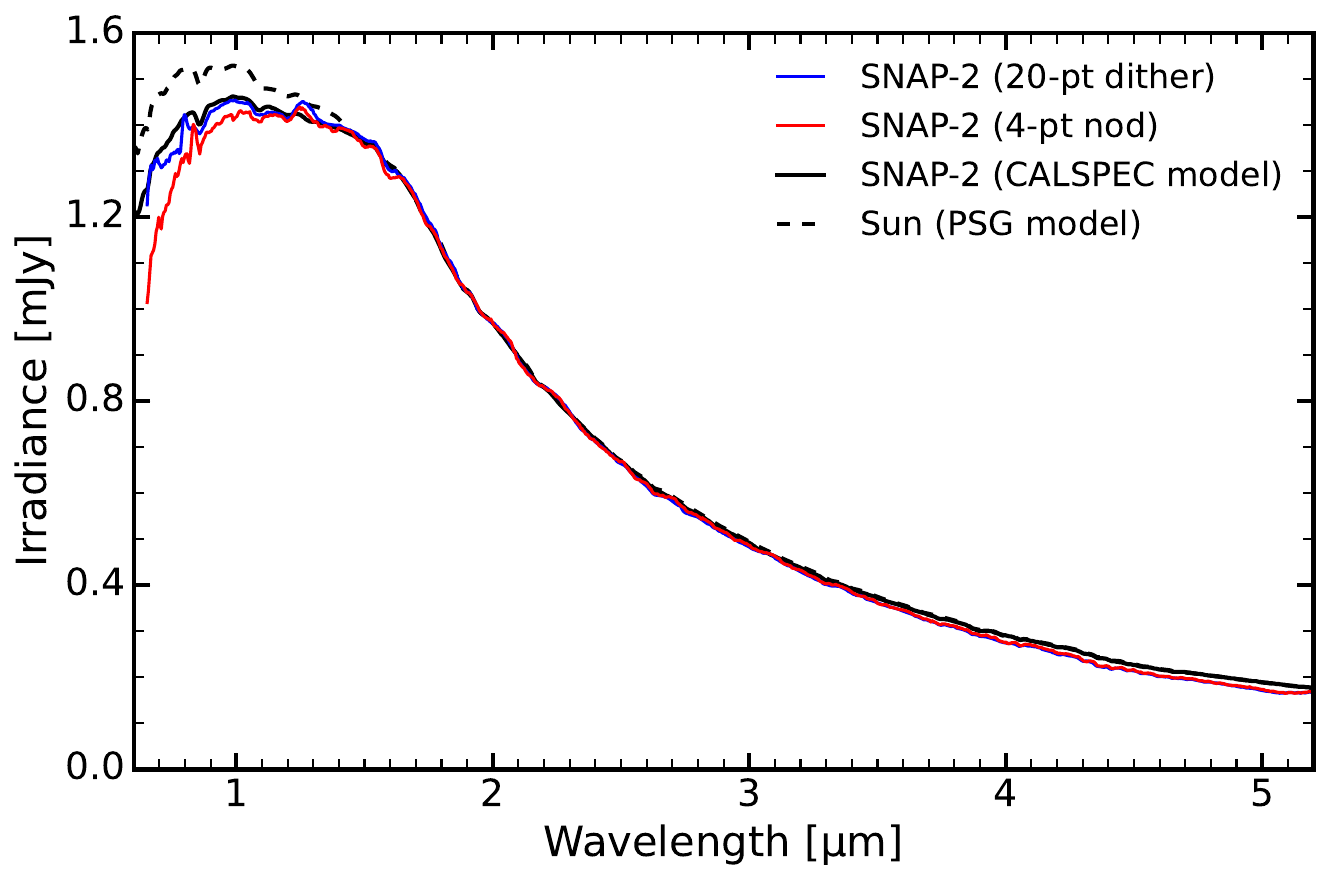}
    \caption{Comparison plot of spectra of the solar analog star SNAP-2 obtained from two JWST observations that utilized different dither patterns (blue and red curves). The CALSPEC model spectrum of SNAP-2 is overplotted in the solid black line. The difference between the two JWST spectra of SNAP-2 at short wavelengths is attributed to variations in sensitivity across the NIRSpec IFU detector that are not corrected by the current calibration pipeline. The solar standard spectrum from the PSG is also included (black dashed line), which shows a discrepant spectral slope at short wavelengths from the SNAP-2 model spectrum. All spectra have been renormalized to match the blue curve at 2.0~\um.}
    \label{fig:stars}
\end{figure}

Figure~\ref{fig:stars} shows the irradiance spectra of SNAP-2 from the two observations, both extracted using a 3-pixel-radius aperture. Also plotted is the CALSPEC model spectrum of SNAP-2 \citep{bohlin2014}. The two NIRSpec spectra have largely equivalent shapes, except at wavelengths shorter than about 1.4~\um: here, the spectrum measured from the 4-point nod observation drops off significantly, indicating a relative decrease in detector sensitivity toward the edges of the FOV. Meanwhile, the spectrum derived from the 20-point dither pattern is in good agreement with the CALSPEC model. It follows that the choice of one or the other SNAP-2 spectrum affects the continuum shape of the resultant reflectance spectrum, specifically the short-wavelength spectral slope.

\subsection{Reflectance Spectra}

During our observation, 2001~XR254 was situated at the nominal pointing position near the center of the FOV, and as such, we expect the wavelength-dependent sensitivity profile of the extracted spectrum to closely mirror that of the SNAP-2 spectrum measured from the 20-point dither observation. On the other hand, the location of 2016~BP81 relative to the center of the FOV in all four dither positions is more similar to the region of the detector sampled by the 4-point nod observation of SNAP-2. We therefore paired the irradiance spectra of 2001~XR254 and 2016~BP81 with the 20-point dither and 4-point nod spectra of SNAP-2, respectively, when deriving the reflectance spectra. 

\begin{figure}[t!]
    \includegraphics[width=\columnwidth]{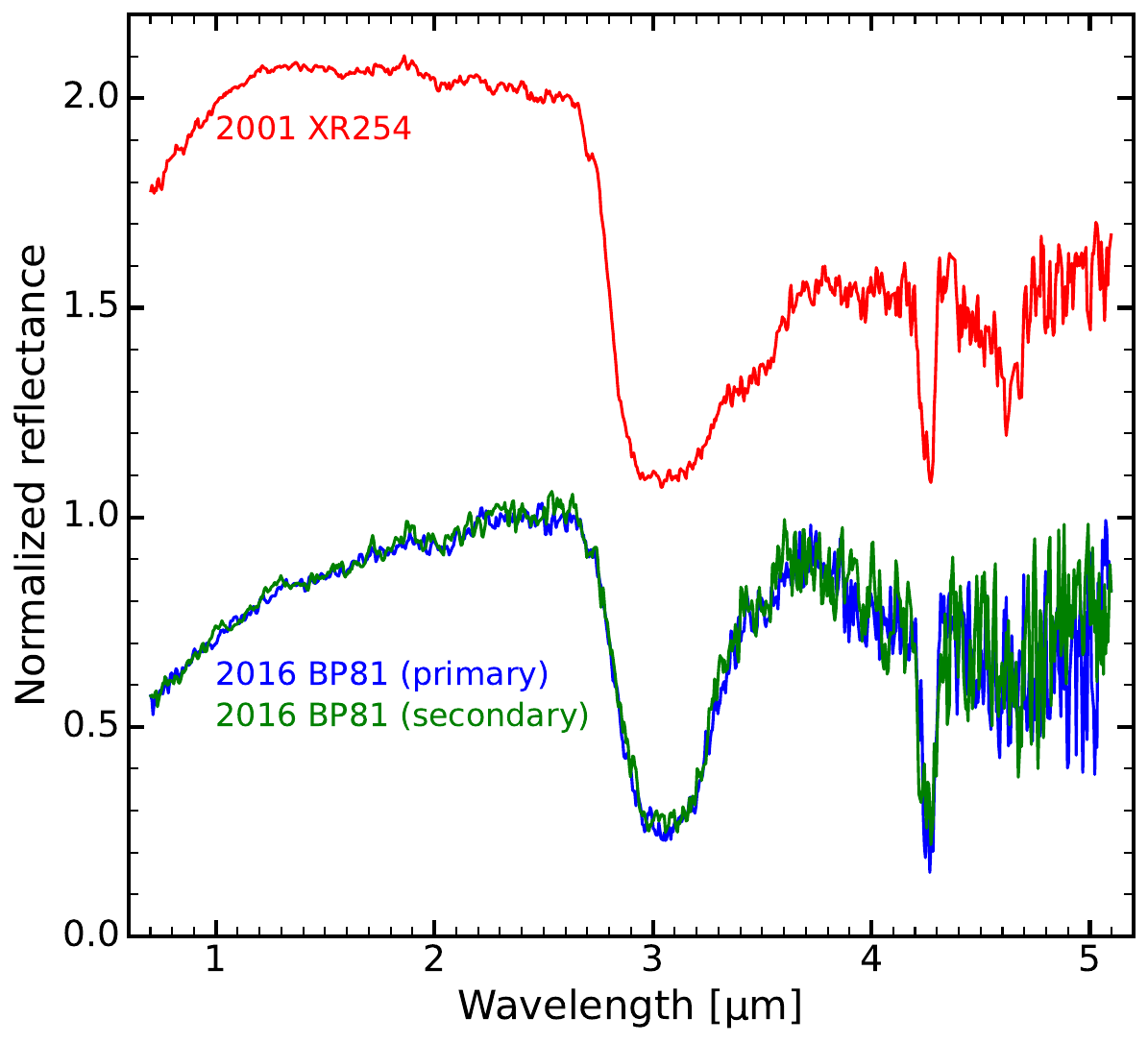}
    \caption{Reflectance spectra of 2001~XR254 and 2016~BP81, normalized to unity at 2.5~\um\ and offset for clarity. The individually extracted spectra of the primary and secondary components of 2016~BP81 are overlaid on top of one another and show identical spectral profiles.}
    \label{fig:reflectance}
\end{figure}

Lastly, we needed to account for the systematic difference between the spectra of SNAP-2 and the Sun. In Figure~\ref{fig:stars}, we include the solar standard spectrum from the Planetary Spectrum Generator (PSG; \citealt{psg}), which has been resampled and convolved to match the wavelength grid and spectral resolution of the NIRSpec PRISM grating. A marked discrepancy in spectral shape is seen at short wavelengths, reflecting a difference in temperature between the two stars. We corrected the reflectance spectra of our targets by multiplying the ratio between the spectra of SNAP-2 and the Sun. Figure~\ref{fig:reflectance} shows the final reflectance spectra of 2001~XR254 and 2016~BP81 (both components), normalized to unity at 2.5~\um.

\subsection{Band-depth Calculations}\label{subsec:depths}
To facilitate quantitative comparisons between spectra, we calculated the depths of the main absorption features of \water~and \cotwo~ice, which lie at 2.0, 3.0, and 4.27~\um. For each band, we stipulated narrow wavelength ranges on one or both sides of the feature that we used to establish the continuum level across the band. We also established a central wavelength region in the middle of the band that we take as the band center. Next, we carried out an error-weighted linear least-squares fit to the points within the continuum region(s) to model the local continuum. This continuum model was then divided away from the reflectance spectrum. 

The band depth was computed by subtracting the weighted average of all points within the band center region from one. The typical flux uncertainties provided by the empirical PSF fitting method in \texttt{jwstspec} (see Section~\ref{subsec:extract}) underestimate the true scatter of the data and cannot account for correlated noise across the wavelength axis. In order to derive more realistic band-depth values and uncertainties, we calculated the ratio between the scatter in the residuals from the continuum fit and the median flux uncertainties within the continuum region(s). The flux uncertainty array of the spectrum was inflated by this ratio when calculating the band depth. The uncertainty in the band depth was set to the standard deviation of the data points within the band center region.

For the 2~\um~\water-ice absorption, the continuum regions were set to 1.80--1.90 and 2.20--2.30~\um, and the band center range was taken to be 2.00--2.05~\um. Only the left side of the 3~\um~\water-ice band (2.60--2.65~\um) was used to establish a constant continuum level across the broad absorption; the band depth was calculated at 2.975--3.025~\um. The \cotwo-ice band depth at 4.26--4.28~\um\ was calculated using the continuum regions 4.00--4.15 and 4.35--4.45~\um.

\section{Discussion}
\label{sec:disc}

\begin{figure}
    \includegraphics[width=\columnwidth]{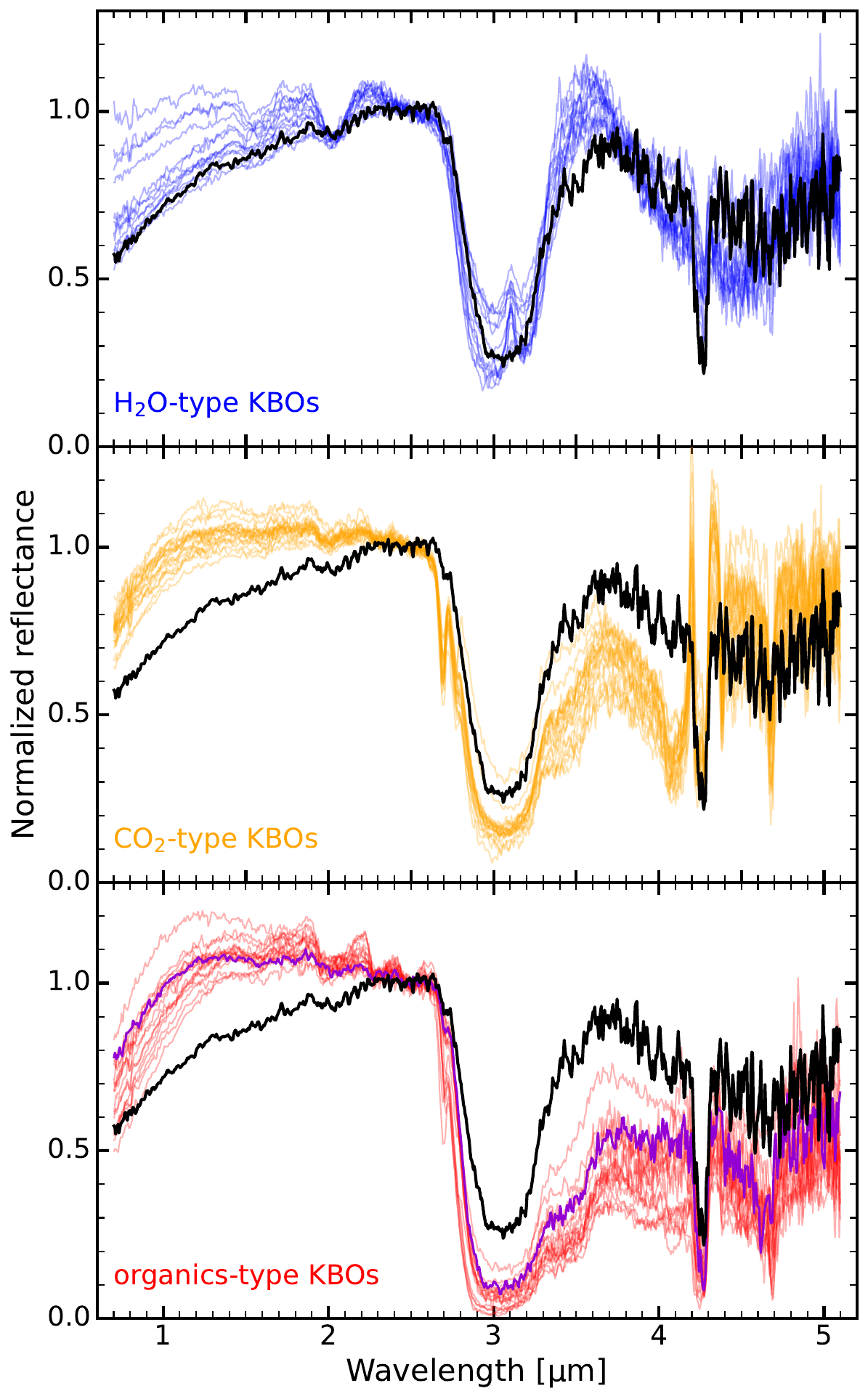}
    \caption{Spectral comparison plot of 2001~XR254 and 2016~BP81 with the KBOs presented in \citet{pinillaalonso2025}. The latter are divided into panels based on their compositional class. All spectra are normalized to unity at 2.5~\um. The spectrum of 2016~BP81, shown in the black curve in each panel, has the greatest affinity to the \water-type KBOs. Meanwhile, 2001~XR254 (purple line in the bottom panel) has a surface composition that is consistent with organics-type objects.}
    \label{fig:discocompare}
\end{figure}

From an initial visual inspection of the reflectance spectra, we find that the primary and secondary components of 2016~BP81 have indistinguishable spectral profiles, with a positive continuum slope throughout the 0.7--2.5~\um\ wavelength range, a broad 3~\um~\water-ice absorption band, an additional weak \water-ice feature centered at 2~\um, a deep primary $\nu_{3}$ \cotwo-ice band at 4.27~\um, and clear evidence for the $\sim$2.7~\um~$\nu_{1}+\nu_{3}$~\cotwo~combination band. The strong agreement between the binary components' spectra indicates that the two bodies have the same surface composition. This finding is in line with models of binary formation that involve gravitational collapse, in which both components accrete out of a single self-gravitating pebble cloud \citep[e.g.,][]{youdin2005,nesvorny2010,nesvorny2019b,parker2011}. Given the identical spectral profiles of the 2016~BP81 binary components, we averaged the two together and utilize the combined spectrum in the following discussion.

Meanwhile, the spectrum 2001~XR254 strongly contrasts with that of 2016~BP81. It is characterized by a much steeper spectral slope below 1.0~\um, which transitions sharply to a negative-sloped continuum through 2.5~\um. At longer wavelengths, the right side of the broad 3~\um~\water-ice feature rises to significantly lower reflectance values than the left side, with a distinct absorption band between 3.3 and 3.6~\um\ characteristic of aliphatic organics. In addition to deep \cotwo\ features, 2001~XR254 also displays a strong CO-ice absorption band centered at 4.7~\um, which is absent on 2016~BP81. Minor features between 2.2 and 2.6~\um\ are suggestive of solid-phase \methanol. Altogether, these distinct properties describe a surface composition very unlike that of the blue binary 2016~BP81 and are instead fully consistent with the organics-type compositional class of KBOs uncovered by \citet{pinillaalonso2025}, all the members of which belong to the red color subpopulation. We therefore classify 2001~XR254 as a red binary cold classical object; indeed, the relatively imprecise measured visible spectral slope of $s=15.2\pm4.4$~\%/100~nm reported in \citet{fraser2021} straddles the threshold between the less-red and red KBO subpopulations.

To examine our two targets more closely within the broader context of known KBO surface types, we reprocessed the NIRSpec observations of KBOs obtained as part of the large Cycle 1 program presented in \citet{pinillaalonso2025}, following an identical methodology to the one we applied to our cold classical targets. We excluded all of the Centaurs (inward-scattered KBOs on orbits within the giant planet region), due to the likelihood of secondary evolution across their surfaces from the enhanced temperatures and space weathering they experience; strong evidence of this trend was recently demonstrated in a dedicated analysis of their JWST spectra \citep{licandro2025}. Likewise, the exceptionally \water-ice-rich Haumea collisional family member 2003~UZ117 was not considered. The size range spanned by this spectroscope sample is 100--800~km, which overlaps at the lower end with the diameters of the cold classical targets from our program. Figure~\ref{fig:discocompare} shows the results of this analysis, with all spectra normalized to unity at 2.5~\um\ and grouped by compositional class. The combined 2016~BP81 spectrum is overplotted in each panel with the black curve, while the spectrum of 2001~XR254 is displayed in purple alongside the other organics-type objects. 

From the figure, it is evident that the reflectance spectra of 2001~XR254 and 2016~BP81 are largely consistent with the range of spectral shapes attested across the organics- and \water-type compositional classes, respectively. In the case of 2001~XR254, all of its spectral properties---the steep spectral slope below 1.0~\um, the negative-sloped continuum between 1.2 and 2.5~\um, the presence of \methanol\ ice features, the overall shape of the broad 3~\um\ absorption, the clear presence of aliphatic organics, the shape of the $\nu_{3}$ \cotwo-ice band, and the profile of the CO-ice feature at 4.7~\um---are archetypal of the organics-type KBOs. In particular, the contribution of \methanol\ in the surface composition is a property that, while not shared by all organics-type KBOs, is only attested among members of that compositional class \citep[][]{brunetto2025}. We also note that the sample from \citet{pinillaalonso2025} includes five other red cold classicals (three of them binaries), all of which are organics-type KBOs, reinforcing our previous statement that 2001~XR254 is a red binary cold classical. Due to our stated focus on blue binaries in this paper, we do not consider 2001~XR254 in the following discussion. 

For 2016~BP81, the overall continuum shape, absence of CO ice, and weak $\nu_{1}+\nu_{3}$ \cotwo\ combination band in its spectrum are broadly consistent with the \water-type KBO class. However, several aspects of its spectrum suggest that the blue binary may represent a distinct type of surface composition that, while generally similar to the \water-type objects shown in Figure~\ref{fig:discocompare}, is somewhat adjacent in the space of spectral properties. It displays a relatively deep 4.27~\um~\cotwo\ band, a very weak 2~\um~\water-ice feature, and lacks the 3.1~\um~\water-ice Fresnel peak. While each of these characteristics is present on other \water-type KBOs in the sample from \citet{pinillaalonso2025}, the combination of all three appears to be unique. Moreover, the reflectance level of the 2016~BP81 spectrum in the vicinity of 3.5~\um\ is substantially lower than the values spanned by the other spectra, suggesting that the surface of 2016~BP81 may be enriched in aliphatic organics relative to the typical \water-type KBO and/or characterized by a different ratio of amorphous vs. crystalline \water\ ice (see the discussion below).

To investigate the relationship between 2016~BP81 and other \water-type KBOs in a more quantitative manner, we carried out a systematic band shape analysis across the \water-type subpopulation. Here, we included some additional \water-type objects from other observing programs: (1) seven Neptune Trojans from Program \#2550 (PI: L. Markwardt; \citealt{markwardt2023}), and (2) three extreme trans-Neptunian objects (ETNOs) from Program \#4665 (PI: B. Holler). All of these datasets were passed through an analogous data reduction and spectral extraction workflow. We measured the band depths of the 2 and 3~\um~\water-ice absorptions, as well as the primary \cotwo\ band at 4.27~\um, following the methodology described in Section~\ref{subsec:depths}.

\begin{figure}
    \includegraphics[width=\columnwidth]{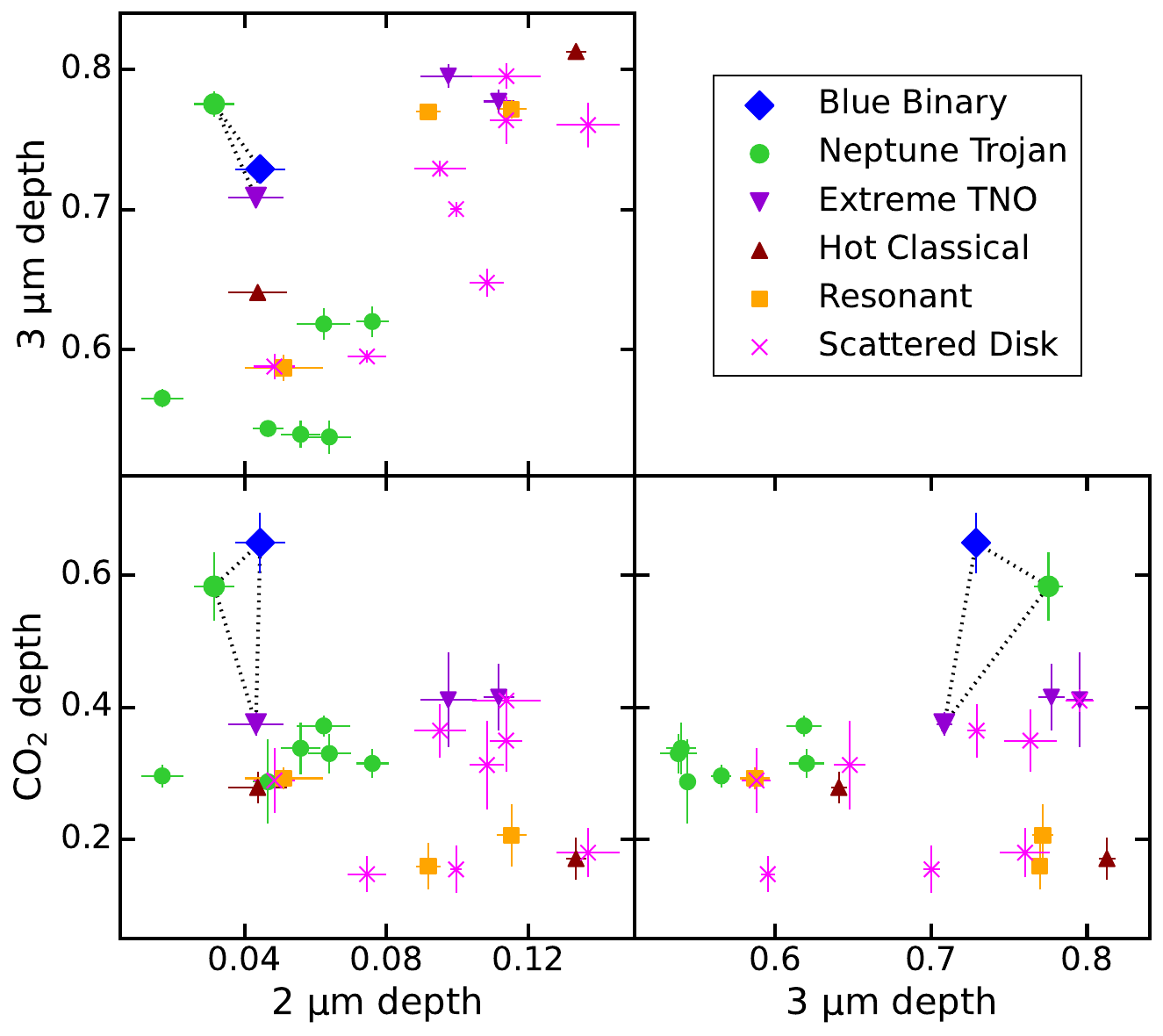}
    \caption{Grid plot of two-parameter distributions for the measured 2~\um~\water, 3~\um~\water, and 4.27~\um~\cotwo~absorption bands among objects in the \water-type KBO compositional class. Different colors and shapes denote the corresponding dynamical classes of the objects. Dashed black lines connect the blue binary 2016~BP81 with the Neptune Trojan 2011~SO277 and the ETNO 2016~QV89. The three objects' spectra show comparable spectral properties, which appear to differ systematically from the rest of the \water-type KBOs in the sample.}
    \label{fig:grid}
\end{figure}

\begin{figure}
    \includegraphics[width=\columnwidth]{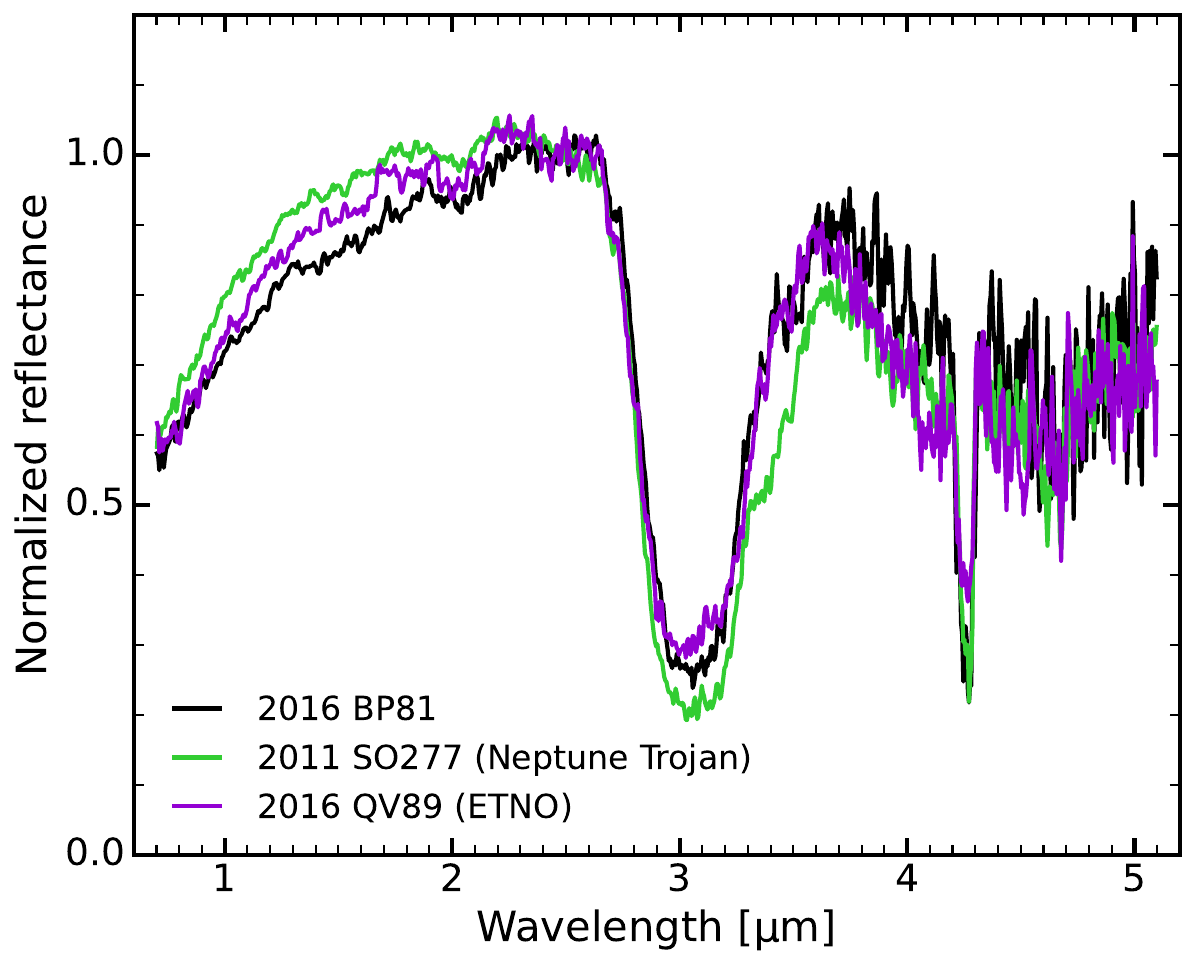}
    \caption{The reflectance spectra of the blue binary 2016~BP81 (black), Neptune Trojan 2011~SO277 (green), and the ETNO 2016~QV89 (purple), normalized to unity at 2.5~\um. All three objects show similar spectral shapes, including weak 2~\um~\water-ice absorption features, deep 3-\um\ \water-ice bands without Fresnel peaks, and strong primary \cotwo-ice bands at 4.27~\um. 2011~SO277 shows an additional distinct absorption band at 3.3--3.6~\um~due to aliphatic organics.}
    \label{fig:indiv}
\end{figure}

The grid of two-parameter distribution plots is shown in Figure~\ref{fig:grid}. Different symbols denote the respective dynamical subpopulations that the targets belong to, with the dynamical classifications taken from the recent work by \citet{volk2024}.

The relationship between the 2 and 3~\um~band depths is an important probe for studying variations in the relative abundance and physical properties of \water~ice across a sample of compositionally similar surfaces. Looking at the top panel in Figure~\ref{fig:grid}, we find that the 2 and 3~\um~\water-ice band depths for most of the \water-type KBOs trace out a positive correlation line. However, 2016~BP81 is visibly offset from this trend and is situated within a tight cluster alongside the Neptune Trojan 2011~SO277 and the ETNO 2016~QV89 at higher 3 vs. 2~\um~depth ratios. The spectra of these three objects are plotted together in Figure~\ref{fig:indiv}, from which the similarity in \water-ice band shapes is evident. The positions of the three objects in the band-depth grid plot are linked by dashed black lines to facilitate interpretation.

Previous laboratory work \citep[e.g.,][]{mastrapa2009,mastrapa2013} and surface modeling of \water-ice-rich solar system minor bodies \citep[e.g.,][]{protopapa2014,raponi2016} have demonstrated that the 2 and 3~\um~band depths scale in tandem when the relative \water-ice abundance on the surface is varied. Meanwhile, changes to the \water-ice grain size modulate the relative depths between the two bands, with a higher 3 vs. 2~\um~depth ratio indicative of smaller grain sizes. While at face value the systematically enhanced 3 vs. 2~\um~band-depth ratios of 2016~BP81, 2011~SO277, and 2016~QV89 suggest surfaces with smaller \water-ice grain sizes than on the other \water-type KBOs, other possible confounding factors warrant caution when considering this interpretation.

Based on laboratory measurements of the wavelength-dependent absorption coefficient of \water\ ice, the 2~\um~band probes significantly deeper penetration depths than the 3~\um~feature \citep[e.g.,][]{mastrapa2013}. It follows that vertical stratification within the uppermost layers of the surface can dramatically alter the relative depths of the \water-ice bands, independent of the size of the ice grains (see, for example, the discussion in \citealt{protopapa2024} related to compositional modeling of the JWST/NIRSpec spectrum of Pluto's satellite Charon). Similarly, the specific configuration in which the \water\ ice is incorporated into the surface regolith---whether as spatially segregated units or within intimate mixtures with other ices and/or opaque materials---has a major impact on the manner in which the various molecular features are expressed in the reflectance spectrum.

Another key physical property of \water~ice is crystallinity. While the mixing ratio between the crystalline and amorphous phases of \water\ ice does not have as large of an impact as grain size on the relative band depths, it does fundamentally set the shape of the 2 and 3~\um~features \citep[e.g.,][]{mastrapa2013,protopapa2014}. In particular, the presence of a prominent Fresnel reflectance peak, as seen on many of the \water-type KBOs plotted in Figure~\ref{fig:discocompare}, is usually interpreted as an indicator of crystalline \water\ ice \citep[e.g.,][]{brown2006,pinillaalonso2025}. It follows that the absence of such a peak in the spectrum of 2016~BP81 (as well as other \water-type objects) suggests a surface where amorphous \water~ice is more dominant. However, given that a sharp Fresnel peak centered at 3.1~\um~requires grains of pure crystalline water ice that are larger than a few microns \cite[e.g.,][]{mastrapa2009,stephan2021}, the strength of this feature may be attenuated due to impurities in the ice grains (e.g., mixtures with other compounds such as \cotwo), very small grain sizes, or vertical stratification. Detailed spectral modeling, which is beyond the scope of this paper, is necessary to robustly disentangle the effects of grain size, crystallinity, mixing, and layering on the \water-ice features in these spectra.

When examining the 3.3--3.6~\um\ wavelength range, we find that the normalized spectrum of 2016~BP81 lies at a much lower reflectance level than the average \water-type KBO. As illustrated in Figure~\ref{fig:indiv}, 2016~QV89 exhibits a similar spectral shape to 2016~BP81 in this region, while 2011~SO277 displays an additional characteristic inflection near 3.3~\um\ that signals the presence of a distinct absorption band due to aliphatic organics, similar to the analogous features seen on \cotwo- and organics-type KBOs (see Figure~\ref{fig:discocompare}). Given that the spectra of these three objects show a tight affinity throughout the rest of the NIRSpec wavelength range, the confirmed signature of organics on 2011~SO277 suggests that the surfaces of all three objects may be systematically enriched in aliphatic organics relative to the other \water-type KBOs. Thus, when specifically considering the abundance of organics, these objects could be interpreted as being transitional between the typical \water-type KBOs and the other two compositional classes. However, the distinct spectral profile of these three objects in this region could also be at least partially explained by a difference in \water-ice crystallinity. Specifically, when controlling for grain size, the curvature and maximum reflectance level of the \water-ice model spectra in the 3.3--3.6~\um\ range are highly sensitive to crystallinity, with the crystalline phase showing systematically lower reflectance levels than the amorphous phase \citep{mastrapa2009,protopapa2014}. In light of this ambiguity in the interpretation of the spectra, we consider the relative enhancement of aliphatic organics on the blue binary 2016~BP81 to be tentative.

When assessing the plots of the \cotwo\ band depth as a function of the 2 and 3~\um~\water~band depths in Figure~\ref{fig:grid}, we once again observe a marked distinction between the blue binary 2016~BP81 and the rest of the \water-type KBOs. While most objects have relative \cotwo\ band depths between 0.1 and 0.4 that are uncorrelated with their 2- or 3-\um\ \water-ice absorption feature depths, both 2016~BP81 and 2011~SO277 have significantly deeper \cotwo-ice spectral signatures. Meanwhile, although 2016~QV89 does not stand out within the broader distribution of \cotwo\ band depths, it nevertheless lies at the upper end of the range spanned by the other \water-type KBOs. From Figure~\ref{fig:indiv}, we can see how comparable the \cotwo\ primary band shapes (and indeed the entire 3.5--5.1~\um\ spectral profiles) are among all three objects. 
 
The discrepant spectroscopic behavior of 2016~BP81, 2011~SO277, and 2016~QV89 points toward a distinct surface subtype that is adjacent to the range spanned by typical \water-type KBOs within the space of spectral properties. We consider differences in current thermal and space weathering environment to be an unlikely explanation for this distinction, given that the three objects have drastically different present-day orbits. Here, 2011~SO277 is particularly diagnostic in this regard, as it lies within the very narrow range of average surface temperatures and energetic particle fluences spanned by the 1:1 mean motion resonance with Neptune while being compositionally distinct from the other six \water-type Neptune Trojans in our sample. Instead, we posit that the distinctive surface properties may reflect differences in the objects' initial formation environment. 

With only one blue binary in the combined spectroscopic sample, it is impossible to make general statements about the blue binary cold classical population. Observations of additional targets are necessary to determine if the distinctive properties of 2016~BP81 are a characteristic trait of the population or if instead the blue binaries include objects with surface properties that are more typical of the broader \water-type KBO class (as is the case with the Neptune Trojans and ETNOs). The answer to that question has fundamental implications for our understanding of the origin and dynamical evolution of the blue binaries. 

\citet{nesvorny2022} put forward two proposals to explain the presence of blue binaries within the cold classical region (see the Introduction), and we now revisit them with the observed properties of 2016~BP81 in mind. In the gentle pushout scenario, dynamical modeling has demonstrated that the survival of the most widely separated blue binaries requires an initial source region beyond 30~au, which necessarily places the primordial less-red/red color transition at even larger heliocentric distances. Meanwhile, the overall implantation efficiency increases with increasing initial orbital semimajor axis, which means that the current blue binary population should preferentially sample the region of the planetesimal disk nearest to the color transition. From the spectrum of 2016~BP81, we do find distinctive spectral features  that could plausibly be interpreted as being transitional between those typical of the less-red \water-type KBOs and the redder \cotwo- and organics-type compositional classes. These include the exceptionally strong \cotwo~primary band and the potentially higher abundance of aliphatic organics, as well as the relatively weak 2~\um~\water-ice absorption (see Figure~\ref{fig:discocompare}).

The pushout hypothesis allows for cold classical implantation from a wide range of initial heliocentric distances within the outer protoplanetary disk \citep[e.g.,][]{nesvorny2020}. Therefore, one may expect future observations of blue binaries to reveal a range of surface compositions, including those that are more consistent with the majority of \water-type KBOs in our spectroscopic sample. Within this framework, 2011~SO277 and 2016~QV89 would also be interpreted as originating from the less-red/red transition region prior to emplacement into their respective present-day dynamical populations. However, we recall that the simulations of the pushout scenario presented in \citet{nesvorny2022} predict significant implantation of less-red singletons, which is inconsistent with our current knowledge of the cold classical population. 

Meanwhile, our spectroscopic sample presents another challenge to the pushout scenario---the apparent absence of \cotwo-type objects within the cold classical population. Under the compositional gradient framework described in \citet{pinillaalonso2025}, which places the initial formation region of the \cotwo-type KBOs in between that of the \water- and organics-type classes, one would expect the outward implantation of the \water-type blue binaries from the inner trans-Neptunian disk to be accompanied by a significant admixture of \cotwo-type objects in the red cold classical population. However, all six red cold classicals with published JWST spectra (including 2001~XR254 from our observations) belong to the organics-type compositional class. In fact, out of all dynamical subpopulations within the Kuiper belt, only the cold classicals appear to lack \cotwo-type objects \citep{depra2025,pinillaalonso2025}. Without future refinements to the dynamical models that address the paucity of less-red singletons and \cotwo-type objects within the cold classical population, the pushout hypothesis appears to be untenable.

The alternative scenario described in \citet{nesvorny2022} is the in situ formation of blue binaries at 42--47~au during the early stages of the protoplanetary disk, when temperatures in the disk midplane were higher and fewer volatile ices (e.g., methane and carbon monoxide) were in the solid phase \citep[e.g.,][]{bitsch2015}. Objects that formed later, after the dispersion of the gaseous component of the disk and subsequent dissipative cooling of the disk midplane, would have accreted more volatile ices onto their surfaces and thereby developed the redder, organics-type surfaces typical of most cold classicals. Crucially, this proposed evolutionary pathway obviates the need to invoke dynamical implantation of the blue binaries into the cold classical region, and as such, the absence of less-red singletons does not present a problem for the model. Moreover, the higher gas-to-dust ratio of the young protoplanetary disk would have facilitated the types of aerodynamic interactions that are conducive to binary formation via streaming instabilities and subsequent gravitational collapse \citep{youdin2005,squire2018,nesvorny2021}. It follows that the in situ formation scenario may be the key to explaining the near-100\% binary fraction of the less-red cold classicals \citep{fraser2017,fraser2021}.

From a compositional perspective, the unique conditions in the outermost reaches of the young protoplanetary disk could explain the distinctive spectral features of 2016~BP81 when compared to other \water-type KBOs, which may have instead formed at a later time and at smaller heliocentric distances---within the warmer inner region of the trans-Neptunian planetesimal disk. If this scenario is correct, one would expect all blue binary cold classicals to display the same characteristic compositional signatures as 2016~BP81. Consequently, the spectrally similar objects 2011~SO277 and 2016~QV89 would likely have initially formed alongside the blue binaries before being scattered from the cold classical region, perhaps due to dynamical stirring of the outer protoplanetary disk during the final phases of Neptune's outward migration. 

\section{Summary}
\label{sec:concl}

We presented JWST/NIRSpec observations of two binary cold classical KBOs. The measured reflectance spectra of the two targets differ significantly: 2001~XR254 belongs to the organics-type KBO compositional class, akin to other red cold classicals, with a steep continuum slope below 1.0~\um, strong signatures of aliphatic organics at 3.3--3.6~\um, deep features of \cotwo~and CO ices, and the distinctive presence of \methanol-ice absorption bands between 2.2 and 2.6~\um; in contrast, the blue binary 2016~BP81 shows spectral features that largely coincide with the \water-type KBO compositional class. However, a detailed comparison of 2016~BP81 against the spectra of other \water-type KBOs revealed some notable differences, including a significantly stronger \cotwo-ice absorption feature at 4.27~\um\ and a higher band-depth ratio between the 3 and 2~\um~\water-ice features. Along with the Neptune Trojan 2011~S0277 and the ETNO 2016~QV89, 2016~BP81 appears to delineate a distinct cluster of objects with spectral properties that are broadly similar to those of a typical \water-type KBO but nonetheless differ in several key ways that may be indicative of a separate formation environment from the rest of the \water-type KBO population.

Observations of additional blue binaries within the cold classical region are necessary to ascertain whether the unique properties of 2016~BP81 are a distinctive characteristic of this enigmatic population. Stepping back to appreciate the bigger picture, the blue binary spectrum described in this work contributes a new piece to the grand puzzle of KBO compositional diversity revealed by previous JWST observations, which have sampled almost the full range of dynamical subpopulations within the Kuiper belt. We are now in a position to leverage the compositional information provided by the body of spectroscopic data to obtain new insights into the detailed formation and evolutionary history of the Kuiper belt. Future studies that incorporate both dynamical and compositional considerations promise to greatly expand our understanding of the various physical and chemical processes that shaped the complex orbital architecture and compositional landscape of the Kuiper belt, with potentially transformative implications for our broader understanding of the outer solar system.

\section*{Acknowledgments}

This work is based on observations made with the NASA/ESA/CSA James Webb Space Telescope. The data were obtained from the Mikulski Archive for Space Telescopes at the Space Telescope Science Institute, which is operated by the Association of Universities for Research in Astronomy, Inc., under NASA contract NAS 5-03127 for JWST. These observations are associated with program \#5940. The JWST data used in this paper can be found in MAST: \dataset[10.17909/1wwx-1c65]{http://dx.doi.org/10.17909/1wwx-1c65}.
\facilities{JWST.}
\software{\texttt{astropy} \citep{astropy2013,astropy2018,astropy2022}}, \texttt{jwst} \citep{jwst}, \texttt{jwstspec} \citep{jwstspec}, \texttt{matplotlib} \citep{matplotlib}, \texttt{numpy} \citep{numpy}, \texttt{scipy} \citep{scipy}.


\bibliography{main.bib}{}
\bibliographystyle{aasjournalv7}

\end{document}